\documentclass[%
reprint,
superscriptaddress,
 amsmath,amssymb,
 aps,
 pra,
]{revtex4-1}

\usepackage{graphicx}
\usepackage{dcolumn}
\usepackage{bm}
\usepackage{longtable}
\usepackage{threeparttablex}
\usepackage{amssymb}
\usepackage{amsthm}
\usepackage{amsmath}
\usepackage{lineno}
\usepackage{hyperref}

\begin{document}

\preprint{APS/PRA}

\title{Study of even-parity Rydberg and autoionizing states of lutetium by laser resonance ionization spectroscopy}

\author{R. Li}
\email{ruohong@triumf.ca}
\affiliation{TRIUMF, Vancouver, BC, V6T 2A3, Canada}%
\author{J. Lassen}
\affiliation{TRIUMF, Vancouver, BC, V6T 2A3, Canada}%
\affiliation{University of Manitoba, Winnipeg, MB, R3T 2N2, Canada}%
\affiliation{Simon Fraser University, Burnaby, BC, V5A 1S6, Canada}%
\author{Z. P. Zhong}
\email{zpzhong@ucas.ac.cn}
\affiliation{School of Physical Sciences, University of Chinese Academy of Sciences, Beijing, 100049, China}
\author{F. D. Jia}
\affiliation{School of Physical Sciences, University of Chinese Academy of Sciences, Beijing, 100049, China}
\author{M. Mostamand}
\affiliation{TRIUMF, Vancouver, BC, V6T 2A3, Canada}%
\affiliation{University of Manitoba, Winnipeg, MB, R3T 2N2, Canada}%
\author{X. K. Li}
\affiliation{School of Physical Sciences, University of Chinese Academy of Sciences, Beijing, 100049, China}
\author{B. B. Reich}
\affiliation{TRIUMF, Vancouver, BC, V6T 2A3, Canada}
\author{A. Teigelh\"ofer}
\affiliation{TRIUMF, Vancouver, BC, V6T 2A3, Canada}%
\affiliation{University of Manitoba, Winnipeg, MB, R3T 2N2, Canada}%
\author{H. Yan}
\affiliation{TRIUMF, Vancouver, BC, V6T 2A3, Canada}

\date{\today}

\begin{abstract}
Multi-step laser resonance ionization spectroscopy of lutetium (Lu) has been performed at TRIUMF's off-line laser ion source test stand. The even-parity Rydberg series $6s^2nd$  $^2D_{3/2}$, $6s^2nd$  $^2D_{5/2}$ and $6s^2ns$  $^2S_{1/2}$ were observed converging to the 6s$^2$ ionization potential. The experimental results has been compared to previous work. 51 levels of Rydberg series $6s^2nd$  $^2D_{5/2}$ and 52 levels of Rydberg series $6s^2ns$  $^2S_{1/2}$ were reported new. Additionally six even-parity autoionization (AI) series converging to Lu ionic states $5d6s$  $^3D_1$ and $5d6s$  $^3D_2$ were observed. The level energies of these AI states were measured. The configurations of the AI states were assigned by relativistic multichannel theory (RMCT) within the framework of multichannel quantum defect theory (MQDT).
\end{abstract}

\pacs{Valid PACS appear here}
\maketitle

\section{Introduction}\label{Introduction}

The need for spectroscopic data of rare earth (RE) elements has been increasing in recent years due to a large number of RE lines observed in stellar spectra from both ground and space based observations, especially in chemically peculiar stars. However due to the complex electronic configuration and dense spectrum, detailed spectroscopic information of most RE elements is not available. As the last element in the lanthanide group lutetium (Lu) has a fully filled $4f$ shell and therefore has a rather simple spectrum compared to most other lanthanides. The first investigation of the Lu optical spectrum was made by Meggers and Scribner in 1930 through the emission spectra in arc and sparks\cite{Meg30}. Soon after King determined the classification of 108 Lu lines by controlling the temperature of the electric furnace\cite{King31}. Using this data, Klinkenberg identified the most important electron configuration in Lu and established the general framework of its low-lying energy levels\cite{Klin54}. Since then a wide frequency range of the Lu spectrum has been studied through traditional absorption spectroscopy via a variety of grating and Fourier transform spectrometers\cite{Bovey53}-\cite{Verg78}. Camus and Tomkins in 1972 observed the first six series of Rydberg states converging to the ionization potential (IP). Based on this data they determined the IP of Lu as 43762.39(10)~cm$^{-1}$\cite{Cam72}. 

The advent of laser technology started a wide application of laser Resonance Ionization Spectroscopy (RIS) in studying atomic structure, especially Rydberg and autoionizing states. First observations of the uranium Rydberg series via RIS were reported by Solarz \textit{et al.} \cite{Solarz76} in 1976. They soon afterward extended the technique to lanthanides and determined the IP of most lanthanides by laser spectroscopy \cite{Worden78}. In 1989 Maeda \textit{et al.}\cite{Maeda89} eventually employed RIS to study the even Rydberg series of Lu I. Four Rydberg series $6s^2ns$ $^2S_{1/2}$, $6s^2nd$ $^2D_{3/2, 5/2}$ and $6s^2ng$ $^2G_{9/2}$ were observed and the IP was determined as 43762.60(10)~cm$^{-1}$, which is still quoted as the most reliable and precise value. Vidolova-Angelova \textit{et al.}\cite{Vid92} investigated radiative lifetimes of the $6s^2nd$ $^2D_{3/2}$ series. A number of studies on even-parity AI states of Lu by RIS were also made \cite{Miller82}-\cite{Ogawa99}. However due to the complexity of AI spectra, the configurations of all those reported states were not assigned and the total angular momentum $J$ values of most states remained undetermined.

TRIUMF applies the RIS technique to deliver isobar-suppressed radioactive ion beams to various nuclear physics experiments \cite{Las05}. An off-line Laser Ion Source test stand (LIS STAND) has been built to develop optimal laser ionization schemes for on-line exotic isotope beam delivery \cite{Lav13}. The investigation of atomic structure of the elements under study is part of this development work. RIS studies on atomic structures of Ga, Ca, Al, Sc, Cd, Y and Sb have been performed at LIS stand in recent years \cite{Li13}-\cite{Li16}. In this work, we applied RIS to study the even-parity Rydberg series of Lu $6s^2nd$ $^2D_{3/2}$, $6s^2nd$ $^2D_{5/2}$ and $6s^2ns$ $^2S_{1/2}$, which converge to the IP. The results are compared with the literature and the level energies of 104 new states are reported. Furthermore six new even-parity AI series converging to Lu ionic states $5d6s$ $^3D_1$ and $5d6s$ $^3D_2$ are reported. The assignment of the configuration of these AI states has been attempted with the aid of RMCT theoretical calculations \cite{lee74,lee89,zou95,huang95,yan96,xia01,xia03}.
\\
\\
\\
\\

\section{Experimental Setup }\label{Setup}

\begin{figure}[!htbp]
\begin{center}
\includegraphics[width=0.4 \textwidth]{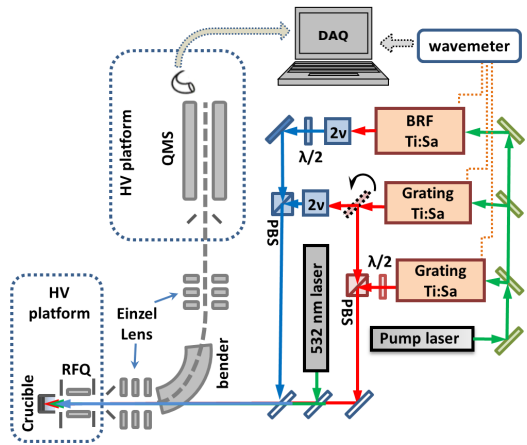}
\end{center}
\centering
\caption{Layout of the TRIUMF resonant ionization laser ion source (TRILIS) test stand consisting of the laser system, vaporization crucible, ion extraction system, quadruple mass spectrometer (QMS) and the PC-based data acquisition (DAQ) system.}
\label{setup}
\end{figure}

The experimental setup is shown in Fig.~\ref{setup}. Three Ti:Sa lasers were employed in this experiment: two grating-tuned and one birefringent-filter-tuned, all pumped by a 50~W 10~kHz pulsed frequency doubled Nd:YAG laser. The Ti:Sa lasers have a typical output power of 1-2~W and linewidth of 1-8~GHz dependent on laser optics, power and wavelength. According to the excitation scheme requirement, the accessible wavelength range could be extended via frequency conversion by employing nonlinear crystals, typically $\beta$-barium borate (BaB$_2$O$_4$) or bismuth borate (BiB$_3$O$_6$) crystals. The laser power after frequency doubling was typically 200-400~mW. The continuously-tunable grating Ti:Sa lasers provided an efficient tool to study atomic structures with the photon energy range of 11000-14000~cm$^{-1}$ \cite{Tei10}. With automated phase matching of a BBO crystal, a continuous scan across the photon energy range of 22300-28200~cm$^{-1}$ was achieved. The Ti:Sa lasers are Q-switched by intracavity Pockel cells to temporally synchronize the laser pulses. The spatial overlap of the multiple laser beams was achieved by polarization beam splitters and dichroic mirrors. For non-resonant ionization, a 4.8~W 532~nm Nd:YVO$_4$ laser (Spectra-Physics Inc.YHP-40) at 10~kHz repetition rate was utilized. The pulse width of the 532~nm laser was $\sim$30~ns. A High Finesse WS/6 wavemeter monitored and measured the laser wavelengths with a precision of 10$^{-6}$. To ensure the accuracy in measurements, the wavemeter was routinely calibrated to a polarization stabilized HeNe laser with a 10$^{-8}$ wavelength accuracy (Melles Griot 05 STP 901/903).

A standard solution (Alfa Aesar Specpure, 1~$\mu$$g$/$\mu$$l$ Lu$_2$O$_3$ in 5\% HNO$_3$ solution) was loaded on a piece of Zr foil. The foil was then dried in an oven at 110$^{\circ}$C, and afterward folded into a small pierce and inserted into a Ta crucible. Lu atomic vapor was generated as the crucible as being resistively heated up to 1500$^{\circ}$C inside the vacuum chamber operated at $\sim$5$\times$10$^{-6}$ Torr. Irradiated by the photons from multiple laser beams, Lu atoms were stepwise excited to high lying Rydberg states and autoionization states. Although having energies below the IP, highly excited Rydberg atoms turned out to ionize due to ambient thermal photons, external electric field or thermal collisions. The generated Lu ions will be guided through a RFQ ion guide and then be extracted and accelerated to 10~keV. This combination of laser ion source (LIS) and RFQ ion guide was named as IG-LIS\cite{Seb14}-\cite{Heg13}. After electrical focusing, the ion beam is deflected 90$^{\circ}$ into a vertically oriented detection system. The detection system consists of a deceleration optics and a quadruple mass spectrometer (EXTREL-QMS MAX300) with an electron multiplier for charged particle detection. A detailed description of the LIS stand can be found in \cite{Lav13}.

\section{Experimental Procedure and Spectroscopy Results of Lutetium }\label{results}

To access different energy regions, a variety of excitation and ionization schemes were chosen (all the wavelengths shown in this paper are the values in air):\\
\\
\resizebox{0.45\textwidth}{!}{
A: $5d6s^2$ $^2D_{3/2}$ $\xrightarrow{356.785~nm}$ $5d6s6p(^1D)$ $^2F^{\circ}_{5/2}$ $\xrightarrow{IR~scan}$ ?$\xrightarrow{532~nm}$}\\
\\
\resizebox{0.4\textwidth}{!}{ B1: $5d6s^2$ $^2D_{3/2}$ $\xrightarrow{337.650~nm}$ $5d6s6p(^1D)$ $^2$$D^{\circ}_{3/2}$ $\xrightarrow{IR~scan}$ ?} \\
\\
\resizebox{0.4\textwidth}{!}{ B2: $5d6s^2$ $^2D_{3/2}$ $\xrightarrow{337.650~nm}$ $5d6s6p(^1D)$ $^2$$D^{\circ}_{3/2}$ $\xrightarrow{blue~scan}$ ?}\\
\\
Using these schemes, three series of even-parity Rydberg states and six series of even-parity AI Rydberg states were observed and measured.


\begin{figure}[!htbp]
\begin{center}
\includegraphics[width=0.5\textwidth]{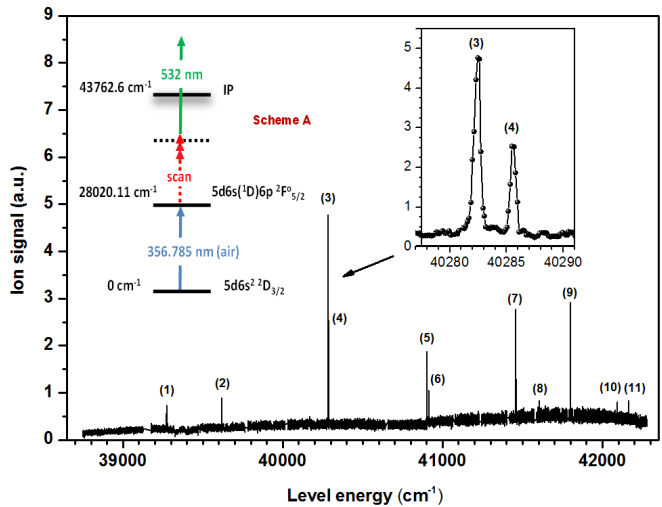}
\end{center}
\centering
\caption{Spectrum of low energy even-parity states of LuI obtained via scheme A. The insert shows the details of two closely spaced resonances (3) and (4).}
\label{A1}
\end{figure}

\subsection{Low energy states via scheme A}

Neutral Lu was excited from the ground state $5d6s^2$ $^2D_{3/2}$ to the excited state of $5d6s6p$ $^2F^{\circ}_{5/2}$ in scheme A. The excitation laser at 356.785~nm  was provided by the frequency-doubled BRF-tuned Ti:Sa laser. The atoms were further excited by a grating-tuned Ti:Sa laser, which allows excitation of Lu atoms to the energy range of 38800-42200~cm$^{-1}$. To finally ionize the excited atoms, a 4.8~W 532~nm Nd:YVO$_4$ laser was employed for nonresonant ionization. The scan was done at a resolution of $\sim$2~GHz/step. Eleven resonances were observed for the second excitation step, as shown in Fig.~\ref{A1}. 

To precisely determine the energies of the resonant levels, fine scans across each resonance were performed with a increased resolution of $\sim$0.2~GHz/step. To eliminate level energy shift in the measurement due to the time delay between the wavelength reading and the ion counting, the scan speed was kept at two data points per second. The measurements at this scan speed showed no systematic shift in central energies of resonances as a function of scan direction. Each resonance was scanned 3-5 times to determine the resonance center energy. The statistical error of the measured level energies is within 0.02~cm$^{-1}$. Three observed resonances have the known upper levels, which were measured by Camus \textit{et al.} in 1972 via absorption spectroscopy \cite{Cam72} and complied into the NIST atomic spectroscopy database (ASD) \cite{NIST}. For these levels, their energies measured in this work agree well with Camus's values within uncertainty (Tab.~\ref{table 337nm scan}).

\begin{table}[!htbp] \footnotesize
\caption{Observed resonant levels excited from the level $5d6s6p$ $^2F^{\circ}_{5/2}$ (scheme A). The statistical error of the measured level energies in this work is within 0.02~cm$^{-1}$. The level energies are compared to the work by Camus \textit{et al.} \cite{Cam72}. }
\begin{center}
\begin{threeparttable}
\resizebox{0.45\textwidth}{!}{
\begin{tabular}{cccccc}
	\hline\hline
&\multicolumn{2}{c}{this work}&&\multicolumn{2}{c}{Camus's \cite{Cam72}}\\
\cline{2-3}\cline{5-6}
label &level energy&$\lambda$  && level energy &level\\
&(cm$^{-1}$)& nm (in vac)&&(cm$^{-1}$) & configuration \\
\hline
(1)& 39272.65&888.688&& & \\
(2)&39616.99&862.301&& & \\
(3)&40282.58&815.497&& & \\
(4)&40285.72&815.288&& & \\
(5)&40901.01&776.343&& 40901.01 &$6s^29d~^2D_{3/2}$\\
(6)&40912.04&775.679&& & \\
(7)&41457.41\tnote{a}&744.197&& & \\
(8)&41605.48&736.086&& 41605.46 &$6s^210d~^2D_{3/2}$\\
(9)&41799.86\tnote{a}&725.703&& & \\
(10)&42092.31&710.621&& 42092.30 &$6s^211d~^2D_{3/2}$\\
(11)&42164.98\tnote{a}&706.970&& & \\
\hline\hline
\end{tabular}}
 \begin{tablenotes}
       \item[a] the levels observed in scheme B1 as well, which constrains their angular momentum $J$ values to 3/2 or 5/2.
 \end{tablenotes}
\end{threeparttable}
\end{center}
\label{table 337nm scan}
\end{table}

\begin{figure*}[!htbp]
\begin{center}
\centerline{\includegraphics[width=0.95\textwidth]{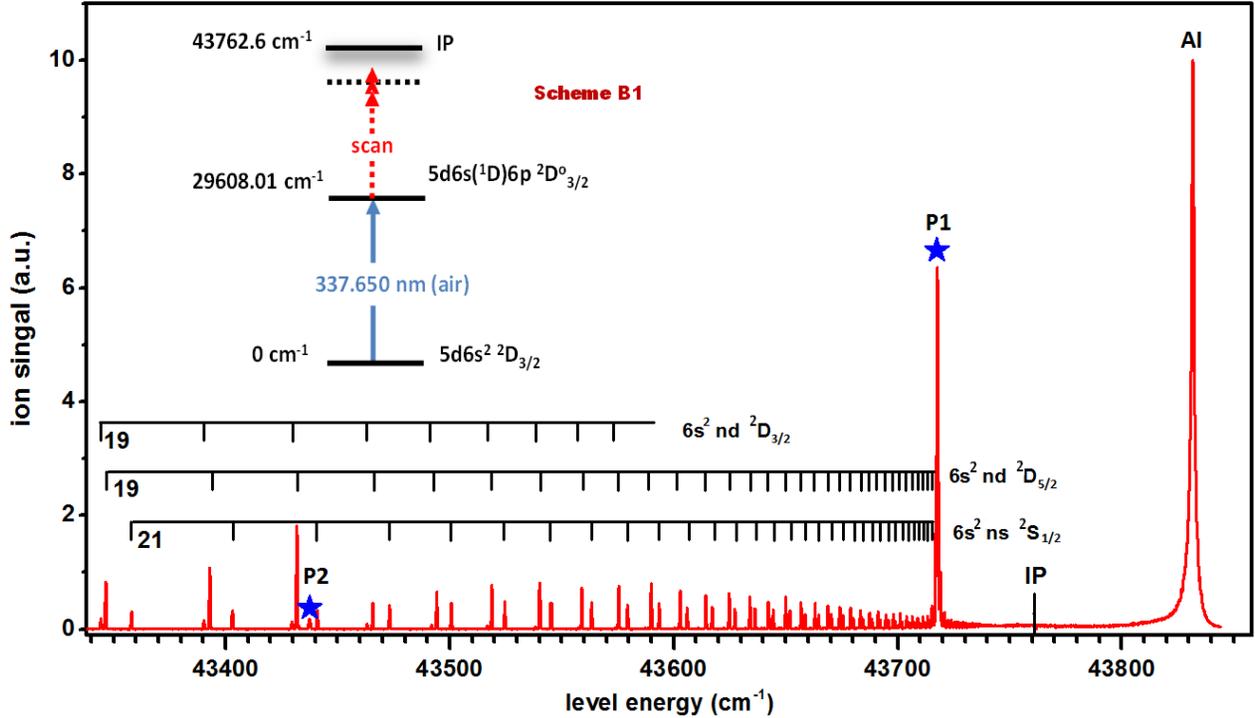}}
\end{center}
\centering
\caption{The Rydberg spectrum of LuI observed via scheme B1. The blue $\bigstar$ show where perturber states P1 and P2 exist.} 
\label{B1}
\end{figure*}

\subsection{Even-parity Rydberg states via scheme B1}\label{B1_Fano}

\begin{figure}[!htbp]
\begin{center}
\centerline{\includegraphics[width=0.5\textwidth]{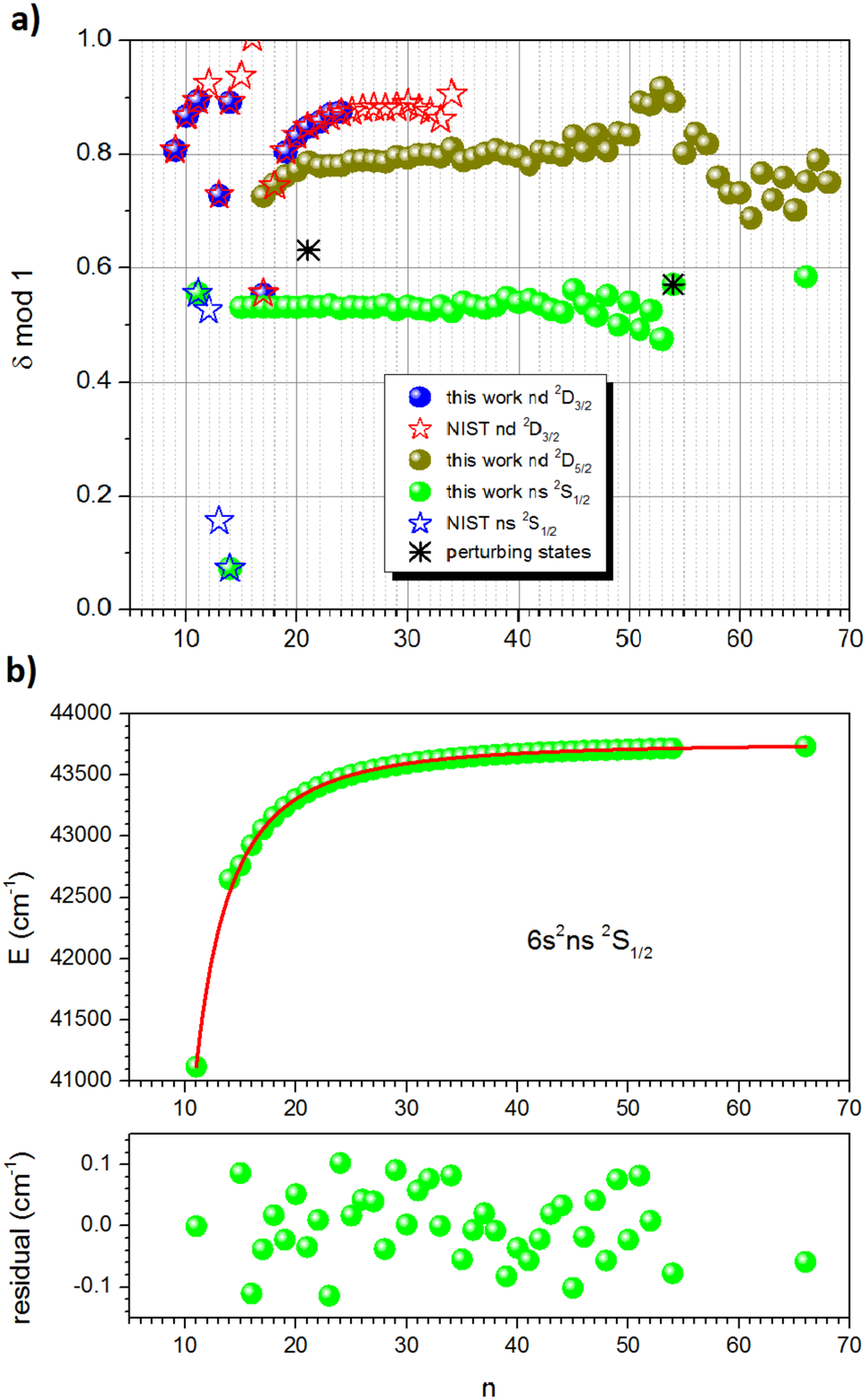}}
\end{center}
\centering
\caption{ a) $\delta$$mod1$ v.s. principal quantum number $n$ for three observed Rydberg series converging to the IP. b) fitting the $6s^2ns$ $^2S_{1/2}$ series to Rydberg-Ritz formula.}
\label{Rydberg}
\end{figure}

To search for higher resonant states, a different first excitation step was used (scheme B1): the BRF laser operated at 337.650~nm was applied to excite Lu atoms from the ground states to the excited state $5d6s6p$ $^2D^{\circ}_{3/2}$ of 29608.01~cm$^{-1}$. Similar to scheme A the second excitation step was provided by the grating-tuned Ti:Sa laser, which can be continuously tuned over 3000~cm$^{-1}$ in the infrared. It allowed the access to the energy range of 40890 - 43870~cm$^{-1}$, which covers the energy region of Lu from Rydberg states, the IP to AI states. The scans were done at $\sim$2~GHz/step. Fig.~\ref{B1} shows a part of the Rydberg spectrum close to the IP. Some observed levels can be easily grouped due to the regularity in the line intensity approaching the IP. However the ambiguity in assignment comes up at low energy end. For clearer classification a Fano plot is used with $\delta$$mod1$ versus $n$ (Fig.~\ref{Rydberg}). Here $\delta$ is the quantum defect and $n$ is the principal quantum number. In the plot the measured levels visually group into three series with $\delta$$mod1$ = 0.88, 0.79 and 0.53, which correspond to Rydberg series of $6s^2nd$ $^2D_{3/2}$, $6s^2nd$ $^2D_{5/2}$ and $6s^2ns$ $^2S_{1/2}$ respectively. The quantum defect values of those series agree with the reported work of Maeda \textit{et al.}\cite{Maeda89}. The identification of low energy members of nd $^2D_{3/2}$ series was assisted by the listed atomic levels in NIST database, which refers to the measurement of Camus \textit{et al.}\cite{Cam72}. The comparisons of this work to Camus's work are presented in both Fig.~\ref{Rydberg}-a and Tab~\ref{table Rydberg_D_3/2}-\ref{table Rydberg_S_1/2}. We also did RMCT calculation for these Rydberg series for comparison. The details of the calculation are given in Sect.~\ref{RMCT}.

Two low energy members of the $ns$ $^2S_{1/2}$ series were measured by adding the 532~nm laser for nonresonant ionization (Tab.~\ref{table Rydberg_S_1/2} footnote). And three low energy states $nd$ $^2D_{3/2}$ $n$=9-11 were measured via scheme A. To our knowledge, there hasn't yet been any report on the level energies of the $6s^2nd$ $^2D_{5/2}$ series for $n$=17-68 and $6s^2ns$ $^2S_{1/2}$ series for $n$=15-66. Maeda \textit{et al.} measured these three series in 1989\cite{Maeda89}. In their published paper, only the determined IP value was reported, but no detailed information about the level energies. The level energies measured in this work are shown in Tab.~\ref{table Rydberg_D_3/2}, ~\ref{table Rydberg_D_5/2}, ~\ref{table Rydberg_S_1/2} for $6s^2nd$ $^2D_{3/2}$, $6s^2nd$ $^2D_{5/2}$ and $6s^2ns$ $^2S_{1/2}$ respectively with the corresponding principal quantum number $n$ and the quantum defect $\delta$. The resonance peaks were scanned at least two times. The peaks at the main Rydberg region ($n$=20-60) were scanned up to six times. The statistical error of the measured level energies is about 0.15~cm$^{-1}$, which includes the uncertainties from the data acquisition delay (estimated by doing the multiple scans in dual directions) and the frequency drift of the first excitation step off the resonance center.

From both the tables and Fig.~\ref{Rydberg}-a, one can see that the quantum defect of the $6s^2ns$ $^2S_{1/2}$ series shows better independence on $n$ in the range of $n$=15-52 compared to other two series, which makes it a better candidate to extracted the IP using the Rydberg-Ritz formula
\begin{equation}
E_{n}=IP-\frac{R_{Lu}}{(n-\delta)^2},
\end{equation}
where $E_n$ is the level energy, $IP$ is the ionization potential, $R_{Lu}$ is the mass-reduced Rydberg constant for Lu, $n$ is the principal quantum number and $\delta$ is the quantum defect. Since there is no obvious $n$-dependence of $\delta$ for $n$=15-52, $\delta$ is treated as a constant in the fitting. The extracted IP value is 43762.52(10)~cm$^{-1}$, which agrees with Maeda's previous measurement 43762.60(10)~cm$^{-1}$\cite{Maeda89}. The value of $\delta$ is 4.5315(6). The uncertainty is the statistical error of the fitted values from six individual scans. An example of the fitted curve and statistics of the residual is shown in Fig~\ref{Rydberg}-b.

\begin{table}[!htbp] \footnotesize
\caption{Even-parity Rydberg series $6s^2nd$ $^2D_{3/2}$ converging to the IP = 43762.6~cm$^{-1}$. } 
\begin{center}
\begin{threeparttable}
\begin{tabular}{ccccccccc}
\\
\hline\hline
 & \multicolumn{5}{c}{this work}&&\multicolumn{2}{c}{Camus'\cite{Cam72}}\\
$n$ & \multicolumn{2}{c}{experimental}&&\multicolumn{2}{c}{RMCT calcualtion}&&\multicolumn{2}{c}{experimental}\\
\cline{2-3}\cline{5-6}\cline{8-9}
 &$\sigma$(cm$^{-1}$)& $\delta$ &&$\sigma$(cm$^{-1}$)& $\delta$ &&$\sigma$(cm$^{-1}$)& $\delta$ \\
\hline
9	&	40901.01\tnote{a}	&	2.81	&		&	40860.38	&	2.85	&		&	40901.01	&	2.81	\\
10	&	41605.48\tnote{a}	&	2.87	&		&	41605.51	&	2.87	&		&	41605.46	&	2.87	\\
11	&	42092.31\tnote{a}	&	2.89	&		&	42101.34	&	2.87	&		&	42092.30	&	2.89	\\
12	&		&		&		&	42444.87	&	2.87	&		&	42430.16	&	2.92	\\
13	&	42722.55	&	2.73	&		&	42692.08	&	2.88	&		&	42722.50	&	2.73	\\
14	&	42873.32	&	2.89	&		&	42875.80	&	2.88	&		&	42873.15	&	2.89	\\
15	&		&		&		&	43016.00	&	2.88	&		&	43008.42	&	2.94	\\
16	&		&		&		&	43125.41	&	2.88	&		&	43112.80	&	3.00	\\
17	&	43236.75	&	2.55	&		&	43212.43	&	2.88	&		&	43236.65	&	2.56	\\
18	&	43290.87	&	2.75	&		&	43282.77	&	2.88	&		&	43291.10	&	2.74	\\
19	&	43344.25	&	2.80	&		&	43340.43	&	2.88	&		&	43344.18	&	2.81	\\
20	&	43390.28	&	2.83	&		&	43388.29	&	2.88	&		&	43390.24	&	2.83	\\
21	&	43429.55	&	2.85	&		&	43428.46	&	2.88	&		&	43429.54	&	2.85	\\
22	&	43463.13	&	2.86	&		&	43462.48	&	2.88	&		&	43463.10	&	2.86	\\
23	&	43491.75	&	2.87	&		&	43491.57	&	2.88	&		&	43491.88	&	2.87	\\
24	&	43516.71	&	2.87	&		&	43516.62	&	2.88	&		&	43516.76	&	2.87	\\
25	&		&		&		&	43538.35	&	2.88	&		&	43538.42	&	2.88	\\
26	&		&		&		&	43557.33	&	2.88	&		&	43557.30	&	2.88	\\
27	&		&		&		&	43573.99	&	2.88	&		&	43573.94	&	2.88	\\
28	&		&		&		&	43588.71	&	2.88	&		&	43588.68	&	2.88	\\
29	&		&		&		&	43601.76	&	2.88	&		&	43601.73	&	2.88	\\
30	&		&		&		&	43613.40	&	2.88	&		&	43613.30	&	2.89	\\
31	&		&		&		&	43623.83	&	2.88	&		&	43623.82	&	2.88	\\
32	&		&		&		&	43633.19	&	2.88	&		&	43633.23	&	2.88	\\
33	&		&		&		&	43641.64	&	2.88	&		&	43641.79	&	2.86	\\
34	&		&		&		&	43649.29	&	2.88	&		&	43649.10	&	2.91	\\

\hline\hline
\end{tabular}
 \begin{tablenotes}
       \item[a] the observed levels from scheme A.
 \end{tablenotes}
   \end{threeparttable}
\end{center}
\label{table Rydberg_D_3/2}
\end{table}

\begin{table}[!htbp] \footnotesize
\caption{Even-parity Rydberg series $6s^2nd$ $^2D_{5/2}$ converging to the IP = 43762.6~cm$^{-1}$. }
\begin{center}
\begin{threeparttable}
\begin{tabular}{cccccc}
\hline\hline
 & \multicolumn{5}{c}{this work}\\
$n$ & \multicolumn{2}{c}{experimental}&&\multicolumn{2}{c}{RMCT calcualtion}\\
\cline{2-3}\cline{5-6}
 &$\sigma$(cm$^{-1}$)& $\delta$ &&$\sigma$(cm$^{-1}$)& $\delta$ \\
\hline
17	&	43223.96	&	2.73	&		&	43234.05	&	2.59	\\
18	&	43290.87	&	2.75	&		&	43300.21	&	2.59	\\
19	&	43346.38	&	2.76	&		&	43354.71	&	2.60	\\
20	&	43392.88	&	2.77	&		&	43400.14	&	2.60	\\
21	&	43431.82	&	2.79	&		&	43438.40	&	2.60	\\
22	&	43465.58	&	2.78	&		&	43470.91	&	2.60	\\
23	&	43494.18	&	2.78	&		&	43498.77	&	2.61	\\
24	&	43518.88	&	2.78	&		&	43522.83	&	2.61	\\
25	&	43540.17	&	2.79	&		&	43543.74	&	2.61	\\
26	&	43558.90	&	2.79	&		&	43562.04	&	2.61	\\
27	&	43575.40	&	2.79	&		&	43578.14	&	2.61	\\
28	&	43589.98	&	2.79	&		&	43592.37	&	2.61	\\
29	&	43602.79	&	2.80	&		&	43605.02	&	2.61	\\
30	&	43614.34	&	2.79	&		&	43616.31	&	2.61	\\
31	&	43624.61	&	2.80	&		&	43626.43	&	2.61	\\
32	&	43633.90	&	2.80	&		&	43635.53	&	2.61	\\
33	&	43642.31	&	2.80	&		&	43643.76	&	2.61	\\
34	&	43649.79	&	2.81	&		&	43651.21	&	2.61	\\
35	&	43656.83	&	2.79	&		&	43657.98	&	2.61	\\
36	&	43663.08	&	2.79	&		&	43664.15	&	2.61	\\
37	&	43668.77	&	2.80	&		&	43669.79	&	2.61	\\
38	&	43673.99	&	2.81	&		&	43674.96	&	2.61	\\
39	&	43678.85	&	2.80	&		&	43679.71	&	2.61	\\
40	&	43683.32	&	2.80	&		&	43684.08	&	2.62	\\
41	&	43687.47	&	2.78	&		&	43688.12	&	2.62	\\
42	&	43691.17	&	2.81	&		&	43691.85	&	2.62	\\
43	&	43694.68	&	2.80	&		&	43695.31	&	2.62	\\
44	&	43697.96	&	2.80	&		&	43698.53	&	2.62	\\
45	&	43700.88	&	2.83	&		&	43701.51	&	2.62	\\
46	&	43703.78	&	2.81	&		&	43704.30	&	2.62	\\
47	&	43706.34	&	2.84	&		&	43706.89	&	2.62	\\
48	&	43708.87	&	2.81	&		&	43709.32	&	2.62	\\
49	&	43711.10	&	2.84	&		&	43711.59	&	2.62	\\
50	&	43713.27	&	2.83	&		&	43713.72	&	2.62	\\
51	&	43715.18	&	2.89	&		&	43715.72	&	2.62	\\
52	&	43717.10	&	2.89	&		&	43717.60	&	2.62	\\
53	&	43718.85	&	2.92	&		&	43719.37	&	2.62	\\
54	&	43720.59	&	2.89	&		&	43721.04	&	2.62	\\
55	&	43722.32	&	2.80	&		&	43722.61	&	2.62	\\
56	&	43723.77	&	2.84	&		&	43724.09	&	2.62	\\
57	&	43725.22	&	2.82	&		&	43725.50	&	2.62	\\
58	&	43726.64	&	2.76	&		&	43726.82	&	2.62	\\
59	&	43727.94	&	2.73	&		&	43728.08	&	2.62	\\
60	&	43729.14	&	2.73	&		&	43729.27	&	2.62	\\
61	&	43730.33	&	2.69	&		&	43730.41	&	2.62	\\
62	&	43731.32	&	2.77	&		&	43731.48	&	2.62	\\
63	&	43732.40	&	2.72	&		&	43732.50	&	2.62	\\
64	&	43733.35	&	2.75	&		&	43733.48	&	2.62	\\
65	&	43734.32	&	2.70	&		&	43734.40	&	2.62	\\
66	&	43735.17	&	2.75	&		&	43735.28	&	2.62	\\
67	&	43735.98	&	2.79	&		&	43736.13	&	2.62	\\
68	&	43736.82	&	2.75	&		&	43736.93	&	2.62	\\
\hline\hline
\\
\\
\\
\end{tabular}
   \end{threeparttable}
\end{center}
\label{table Rydberg_D_5/2}
\end{table}

With their energies below the IP, the Rydberg atoms were eventually ionized by collisions, ambient thermal photons and/or electrical fields. The ionization probability increases as the Rydberg state energies approach the IP. This trend can be readily seen in the intensity distribution of the ion signal at the low energy side of Fig.~\ref{B1}. However when Rydberg state energies further approach to the IP, the ion signal start to drop. This is due to the rapid decrease in the photo-excitation probability to Rydberg states, which scales as $n$$^{*-3}$\cite{Gallagher}. A significant abnormality shows around $52d$ $^2D_{5/2}$ and $54s$ $^2S_{1/2}$, where the ion signal dramatically increases and forms a sharp peak on the ion intensity envelope (marked with a blue $\bigstar$ and P1 in Fig.~\ref{B1}). This normally implies perturbations. The same perturbation was also observed in Maeda's work\cite{Maeda89} and further investigated by microwave spectroscopy\cite{Maeda93}. It was well explained as an Fano interference effect between a doubly excited valence state and a Rydberg series.

\begin{table}[!htbp] \footnotesize
\caption{Even-parity Rydberg series $6s^2ns$ $^2S_{1/2}$ converging to the IP = 43762.6~cm$^{-1}$. }
\begin{center}
\begin{threeparttable}
\begin{tabular}{ccccccccc}
\hline\hline
 & \multicolumn{5}{c}{this work}&&\multicolumn{2}{c}{Camus'\cite{Cam72}}\\
$n$ & \multicolumn{2}{c}{experimental}&&\multicolumn{2}{c}{RMCT calculation}&&\multicolumn{2}{c}{experimental}\\
\cline{2-3}\cline{5-6}\cline{8-9}
 &$\sigma$(cm$^{-1}$)&$\delta$&&$\sigma$(cm$^{-1}$)&$\delta$&&$\sigma$(cm$^{-1}$)&$\delta$\\
\hline
11	&	41120.32\tnote{a}	&	4.56	&		&	40893.74	&	4.82	&		&	41120.27	&	4.56	\\
12	&	41799.89\tnote{a}	&	4.52	&		&	41678.15	&	4.74	&		&	41798.10	&	4.53	\\
13	&		&		&		&	42169.84	&	4.70	&		&	42359.48	&	4.16	\\
14	&	42649.02\tnote{a}	&	4.07	&		&	42502.16	&	4.67	&		&	42649.05	&	4.07	\\
15	&	42761.37	&	4.53	&		&	42738.65	&	4.65	&		&		&		\\
16	&	42928.22	&	4.53	&		&	42913.49	&	4.63	&		&		&		\\
17	&	43056.74	&	4.53	&		&	43046.66	&	4.62	&		&		&		\\
18	&	43157.69	&	4.53	&		&	43150.54	&	4.61	&		&		&		\\
19	&	43238.36	&	4.53	&		&	43233.21	&	4.60	&		&		&		\\
20	&	43304.01	&	4.53	&		&	43300.11	&	4.60	&		&		&		\\
21	&	43357.92	&	4.53	&		&	43355.03	&	4.59	&		&		&		\\
22	&	43402.96	&	4.53	&		&	43400.68	&	4.59	&		&		&		\\
23	&	43440.72	&	4.54	&		&	43439.05	&	4.58	&		&		&		\\
24	&	43473.14	&	4.53	&		&	43471.61	&	4.58	&		&		&		\\
25	&	43500.65	&	4.53	&		&	43499.47	&	4.58	&		&		&		\\
26	&	43524.52	&	4.53	&		&	43523.52	&	4.58	&		&		&		\\
27	&	43545.24	&	4.53	&		&	43544.40	&	4.57	&		&		&		\\
28	&	43563.29	&	4.54	&		&	43562.66	&	4.57	&		&		&		\\
29	&	43579.38	&	4.53	&		&	43578.72	&	4.57	&		&		&		\\
30	&	43593.40	&	4.53	&		&	43592.91	&	4.57	&		&		&		\\
31	&	43606.00	&	4.53	&		&	43605.52	&	4.57	&		&		&		\\
32	&	43617.22	&	4.53	&		&	43616.78	&	4.57	&		&		&		\\
33	&	43627.18	&	4.53	&		&	43626.86	&	4.57	&		&		&		\\
34	&	43636.30	&	4.52	&		&	43635.94	&	4.57	&		&		&		\\
35	&	43644.32	&	4.54	&		&	43644.13	&	4.56	&		&		&		\\
36	&	43651.77	&	4.53	&		&	43651.55	&	4.56	&		&		&		\\
37	&	43658.52	&	4.53	&		&	43658.30	&	4.56	&		&		&		\\
38	&	43664.62	&	4.53	&		&	43664.45	&	4.56	&		&		&		\\
39	&	43670.15	&	4.55	&		&	43670.07	&	4.56	&		&		&		\\
40	&	43675.33	&	4.54	&		&	43675.22	&	4.56	&		&		&		\\
41	&	43680.03	&	4.54	&		&	43679.95	&	4.56	&		&		&		\\
42	&	43684.41	&	4.54	&		&	43684.31	&	4.56	&		&		&		\\
43	&	43688.46	&	4.53	&		&	43688.33	&	4.56	&		&		&		\\
44	&	43692.19	&	4.52	&		&	43692.05	&	4.56	&		&		&		\\
45	&	43695.49	&	4.56	&		&	43695.50	&	4.56	&		&		&		\\
46	&	43698.77	&	4.54	&		&	43698.70	&	4.56	&		&		&		\\
47	&	43701.80	&	4.52	&		&	43701.68	&	4.56	&		&		&		\\
48	&	43704.47	&	4.55	&		&	43704.45	&	4.56	&		&		&		\\
49	&	43707.18	&	4.50	&		&	43707.04	&	4.56	&		&		&		\\
50	&	43709.50	&	4.54	&		&	43709.46	&	4.56	&		&		&		\\
51	&	43711.87	&	4.49	&		&	43711.72	&	4.56	&		&		&		\\
52	&	43713.91	&	4.53	&		&	43713.84	&	4.56	&		&		&		\\
53	&	43715.99	&	4.48	&		&	43715.84	&	4.56	&		&		&		\\
54	&	43717.68	&	4.57	&		&	43717.71	&	4.56	&		&		&		\\
55	&		&		&		&	43719.47	&	4.56	&		&		&		\\
56	&		&		&		&	43721.13	&	4.56	&		&		&		\\
57	&		&		&		&	43722.70	&	4.56	&		&		&		\\
58	&		&		&		&	43724.18	&	4.56	&		&		&		\\
59	&		&		&		&	43725.58	&	4.56	&		&		&		\\
60	&		&		&		&	43726.90	&	4.56	&		&		&		\\
61	&		&		&		&	43728.15	&	4.56	&		&		&		\\
62	&		&		&		&	43729.34	&	4.56	&		&		&		\\
63	&		&		&		&	43730.47	&	4.56	&		&		&		\\
64	&		&		&		&	43731.54	&	4.56	&		&		&		\\
65	&		&		&		&	43732.56	&	4.56	&		&		&		\\
66	&	43733.51	&	4.58	&		&	43733.53	&	4.56	&		&		&		\\
\hline\hline
\end{tabular}
\begin{tablenotes}
       \item[a] the observed levels with the 532nm laser added on scheme B1.
 \end{tablenotes}
   \end{threeparttable}
\end{center}
\label{table Rydberg_S_1/2}
\end{table}

The interference between a valence state and the continuum is known to cause the typical Fano profiles for AI states. The profile can be described by the Fano formula provided only two channels are involved:
\begin{equation}
I(\epsilon)=I_{res}\frac{(q+\epsilon)^2}{(1+\epsilon^2)}+I_{cont}\quad with\quad \epsilon=\frac{E-E_{res}}{\Gamma/2}  
\label{Fano}
\end{equation}
$E_{res}$ is the AI resonance energy and $\Gamma$ is the natural width of the resonance. $I_{cont}$ is the ion signal generated by the interaction with the continuum. The width of the profile $\Gamma$ is proportional (by a factor of 2$\pi$) to the strength of the configuration interaction between the valence state and the continuum, and $\pi$$q$${^2}$/2 is the ratio of the transition probability to the AI resonance and to the continuum in an energy band $\Gamma$ \cite{Gallagher83}. This Fano profile is obviously visible in the AI resonance at 43831.6~cm$^{-1}$ (labeled as AI in Fig.~\ref{B1}), which has been reported by several works previously \cite{Miller82}-\cite{Kujirai98}. The relatively broad linewidth of the state indicates rapid autoionization through strong interaction with the continuum. Meanwhile for this AI state, $\lvert$$q$$\lvert$$\gg$1, which shows a bigger transition probability from the intermediate excited state to the valence states compared to the continuum.

The same theory can be applied to explain the abnormalities on the ion line intensity of Rydberg series caused by perturber states, when treating the Rydberg series as a quasi-continuum. The perturbation at $54s$ $^2S_{1/2}$ was determined to be very weak in Maeda's work\cite{Maeda89}, which is also conspicuous in our data as a very localized change in the ion intensity (at P1 in Fig.~\ref{B1}) in and very small deviation in the quantum defect of the Rydberg series in the vicinity of the perturber. The difference between our and Maeda's experimental observations is that we observed a high ion intensity peak around the perturbation in contrary to an intensity dip observed by Maeda. It most likely came from the delayed ionization (of a few $\mu$s) applied in Maeda's experiment. Due to the admixture of the valence state, the radiative lifetime of the perturbed Rydberg state $54s$ $^2S_{1/2}$ is significantly decreased, which increases the photoexcitation rate to this state. In our experiment, it increased the laser excitation efficiency. However in Maeda's experiment, this lifetime-reduced state most likely could not survive until the arriving of the electric-field pulse for ionization detection.

Another fact to consider is that in our work the core configuration of the intermediate state $5d6s6p(^1D)$ $^2D^{\circ}_{3/2}$ is different from that of the Rydberg states converging to the IP $6s^2$ $^1S_{0}$. It is reasonable to suspect that the perturber state is a low energy member of a series converging to one of the Lu${^+}$ 5d6s ionic states. Due to the same core configuration with respect to the intermediate state, the transition probability from the intermediate state to this perturber state shall be much higher than that to the Rydberg series. Therefore the absolute value of the Fano profile parameter $q$ is much higher than 1, which makes the interference feature appears as a Lorentzian. 


Contrary to the localized and Lorentzian profile at P1, a broad interference feature of $q$$\sim$-1 appears around $21d$ $^2D_{5/2}$, which shows significant interaction with multiple numbers of the Rydberg series. The perturber state here shall has a photoexcitation rate from the intermediate state comparable to that from the Rydberg series it interacts with. Since it interacts with $nd$ $^2D_{5/2}$ series, the perturber state shall have the same total angular momentum number of $J$=5/2. Unlike the perturber at P1 which is very close in energy to a perturbed state, this perturber can be easily resolved at 43437.36~cm$^{-1}$ (marked as a blue $\bigstar$ and P2 in Fig.~\ref{B1}). Due to this big energy difference from the perturber, the quantum defect of $nd$ $^2D_{5/2}$ series does not show any significant deviation in the vicinity of the perturber state. 

\begin{figure}[!htbp]
\begin{center}
\centerline{\includegraphics[width=0.5\textwidth]{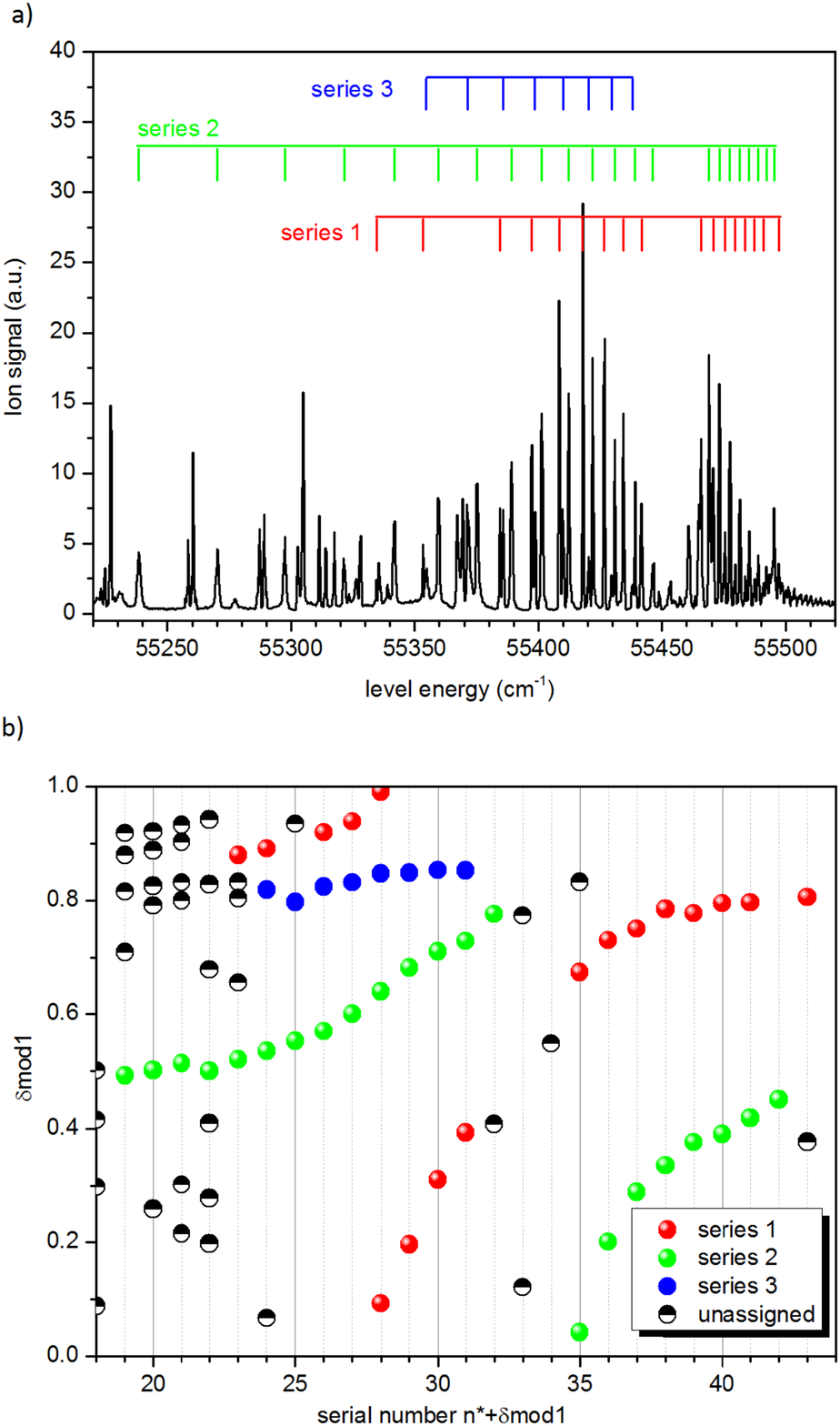}}
\end{center}
\centering
\caption{ a) AI spectrum approaching the limit $5d6s$ $^3D_1$ = 55558.8~cm$^{-1}$. b) $\delta$$mod1$ v.s. serial number ($n$$^*$+$\delta$$mod1$) for three observed AI Rydberg series. The levels are unassigned for the serial number range 32-35 due to the strong perturbation at the energy around 55450~cm$^{-1}$.}
\label{1st_ionic}
\end{figure}

\subsection{Even-parity AI Rydberg states via scheme B2}

With scheme B1, we could only access the energy region close to the IP. To study more AI states, scheme B2 was applied. The first excitation step of scheme B2 is identical to that of scheme B1. But the second excitation step was provided by a frequency doubled Ti:Sa laser. The automated phase matching of the BBO crystal with the wavelength of the grating Ti:Sa laser allows a continuous frequency scan on the second harmonic of the laser photons. With this scheme, the even-parity AI states in the energy range of 51900 - 57800~cm$^{-1}$ with possible transitions from $5d6s6p$ $^2D^{\circ}_{3/2}$ were studied. 

\LTcapwidth=0.5\textwidth

\begin{ThreePartTable}
\begin{longtable}{cccccc}
\caption{Even-parity AI Rydberg series converging to Lu$^+$ $5d6s$ $^3D_1$ = 55558.8~cm$^{-1}$ with both the experimental values and the RMCT calculated values in the assigned configurations.}\\
\label{Rydberg_AI_1}
\begin{TableNotes}
\footnotesize
 \item[*] same serial number for two resonances due to a jump in order.
\end{TableNotes}
\\
\hline\hline
& \multicolumn{2}{c}{{\bf experimental}}&& \multicolumn{2}{c}{{\bf RMCT calculation}}\\
serial& \multicolumn{2}{c}{{\bf Series 1 }}&& \multicolumn{2}{c}{5d6s($^3$D$_1$)ns$_{1/2}$}\\
num.& \multicolumn{2}{c}{}&& \multicolumn{2}{c}{J=1/2}\\
\cline{2-3}\cline{5-6}
&	$\sigma$(cm$^{-1}$)	&$\delta$$mod1$& &$\sigma$(cm$^{-1}$)&$\delta$$mod1$\\
\hline
23	&	55334.57	&	0.88	&		&	55336.79	&	0.77	\\
24	&	55353.35	&	0.89	&		&	55355.52	&	0.77	\\
25	&		&		&		&		&		\\
26	&	55384.39	&	0.92	&		&	55384.36	&	0.92	\\
27	&	55397.27	&	0.94	&		&	55395.93	&	0.05	\\
28\tnote{*}&	55408.41	&	0.99	&		&	55407.42	&	0.08	\\
28\tnote{*}&	55417.94	&	0.09	&		&	55416.59	&	0.23	\\
29	&	55426.58	&	0.20	&		&	55425.25	&	0.34	\\
30	&	55434.35	&	0.31	&		&	55433.79	&	0.38	\\
31	&	55441.70	&	0.39	&		&	55441.62	&	0.40	\\
32	&		&		&		&		&		\\
33	&		&		&		&		&		\\
34	&		&		&		&		&		\\
35	&	55465.71	&	0.67	&		&	55466.12	&	0.60	\\
36	&	55470.62	&	0.73	&		&	55471.18	&	0.62	\\
37	&	55475.33	&	0.75	&		&	55475.79	&	0.65	\\
38	&	55479.61	&	0.78	&		&	55479.8	&	0.74	\\
39	&	55483.73	&	0.78	&		&	55484.48	&	0.58	\\
40	&	55487.44	&	0.80	&		&	55488.11	&	0.61	\\
41	&	55490.95	&	0.80	&		&	55491.43	&	0.65	\\
42	&		&		&		&		&		\\
43	&	55497.20	&	0.81	&		&	55497.44	&	0.72	\\
\hline\hline
\insertTableNotes
\end{longtable}
\end{ThreePartTable}

\begin{longtable}{ccccccccc}
\hline\hline
& \multicolumn{2}{c}{{\bf experimental}}&& \multicolumn{5}{c}{{\bf RMCT calculation}}\\
serial& \multicolumn{2}{c}{{\bf Series 2 }}&& \multicolumn{2}{c}{5d6s($^3$D$_1$)ns$_{1/2}$}&&\multicolumn{2}{c}{5d6s($^3$D$_1$)nd$_{3/2}$}\\
num.& \multicolumn{2}{c}{}&& \multicolumn{2}{c}{J=3/2}&&\multicolumn{2}{c}{J=1/2}\\
\cline{2-3}\cline{5-6}\cline{8-9}
&	$\sigma$(cm$^{-1}$)	&$\delta$$mod1$& &$\sigma$(cm$^{-1}$)&$\delta$$mod1$& &$\sigma$(cm$^{-1}$)&$\delta$$mod1$\\
\hline
19	&	55238.46	&	0.49	&		&	55222.53	&	0.94	&		&	55242.49	&	0.38	\\
20	&	55270.19	&	0.50	&		&	55275.58	&	0.32	&		&	55269.59	&	0.52	\\
21	&	55297.38	&	0.51	&		&	55299.40	&	0.43	&		&	55301.11	&	0.37	\\
22	&	55321.44	&	0.50	&		&	55321.21	&	0.51	&		&	55323.81	&	0.39	\\
23	&	55341.68	&	0.52	&		&	55340.50	&	0.58	&		&	55344.56	&	0.37	\\
24	&	55359.53	&	0.54	&		&	55378.80	&	0.31	&		&	55362.24	&	0.37	\\
25	&	55375.22	&	0.55	&		&	55384.20	&	0.93	&		&	55378.01	&	0.37	\\
26	&	55389.15	&	0.57	&		&	55395.92	&	0.05	&		&	55391.8	&	0.37	\\
27	&	55401.38	&	0.60	&		&	55403.29	&	0.44	&		&	55404.05	&	0.37	\\
28	&	55412.25	&	0.64	&		&	55413.80	&	0.49	&		&	55414.99	&	0.38	\\
29	&	55422.00	&	0.68	&		&	55423.50	&	0.53	&		&	55424.84	&	0.38	\\
30	&	55430.92	&	0.71	&		&	55427.66	&	0.08	&		&	55433.81	&	0.37	\\
31	&	55439.09	&	0.73	&		&	55436.15	&	0.09	&		&	55441.71	&	0.39	\\
32	&	55446.28	&	0.78	&		&	55443.75	&	0.12	&		&	55449.08	&	0.38	\\
33	&		&		&		&		&		&		&		&		\\
34	&		&		&		&		&		&		&		&		\\
35	&	55469.04	&	0.04	&		&	55468.22	&	0.20	&		&	55469.22	&	0.01	\\
36	&	55473.21	&	0.20	&		&	55473.12	&	0.22	&		&	55472.36	&	0.38	\\
37	&	55477.42	&	0.29	&		&	55477.61	&	0.24	&		&	55477.03	&	0.38	\\
38	&	55481.48	&	0.34	&		&	55481.80	&	0.26	&		&	55481.32	&	0.38	\\
39	&	55485.28	&	0.38	&		&	55485.54	&	0.31	&		&	55485.27	&	0.38	\\
40	&	55488.90	&	0.39	&		&	55489.07	&	0.34	&		&	55488.9	&	0.39	\\
41	&	55492.20	&	0.42	&		&	55492.29	&	0.39	&		&	55492.32	&	0.38	\\
42	&	55495.27	&	0.45	&		&	55495.31	&	0.44	&		&	55495.91	&	0.24	\\
\hline\hline
\end{longtable}

\begin{longtable}{cccccc}
\hline\hline
 &\multicolumn{2}{c}{{\bf experimental}}&& \multicolumn{2}{c}{{\bf RMCT calculation}}\\
serial& \multicolumn{2}{c}{{\bf Series 3 }}&& \multicolumn{2}{c}{5d6s($^3$D$_1$)nd$_{5/2}$}\\
num.& \multicolumn{2}{c}{}&& \multicolumn{2}{c}{J=3/2}\\
\cline{2-3}\cline{5-6}
&	$\sigma$(cm$^{-1}$)	&$\delta$$mod1$& &$\sigma$(cm$^{-1}$)&$\delta$$mod1$\\
\hline
24	&	55354.63	&	0.82	&		&	55355.30	&	0.78	\\
25	&	55371.51	&	0.80	&		&	55370.96	&	0.83	\\
26	&	55385.71	&	0.82	&		&	55384.29	&	0.93	\\
27	&	55398.59	&	0.83	&		&	55395.91	&	0.05	\\
28	&	55409.99	&	0.85	&		&	55411.77	&	0.68	\\
29	&	55420.37	&	0.85	&		&	55418.78	&	0.01	\\
30	&	55429.67	&	0.85	&		&	55425.39	&	0.32	\\
31	&	55438.10	&	0.85	&		&	55440.50	&	0.55	\\
\hline\hline
\end{longtable}

\addtocounter{table}{-2}

\begin{figure}[!htbp]
\begin{center}
\centerline{\includegraphics[width=0.5\textwidth]{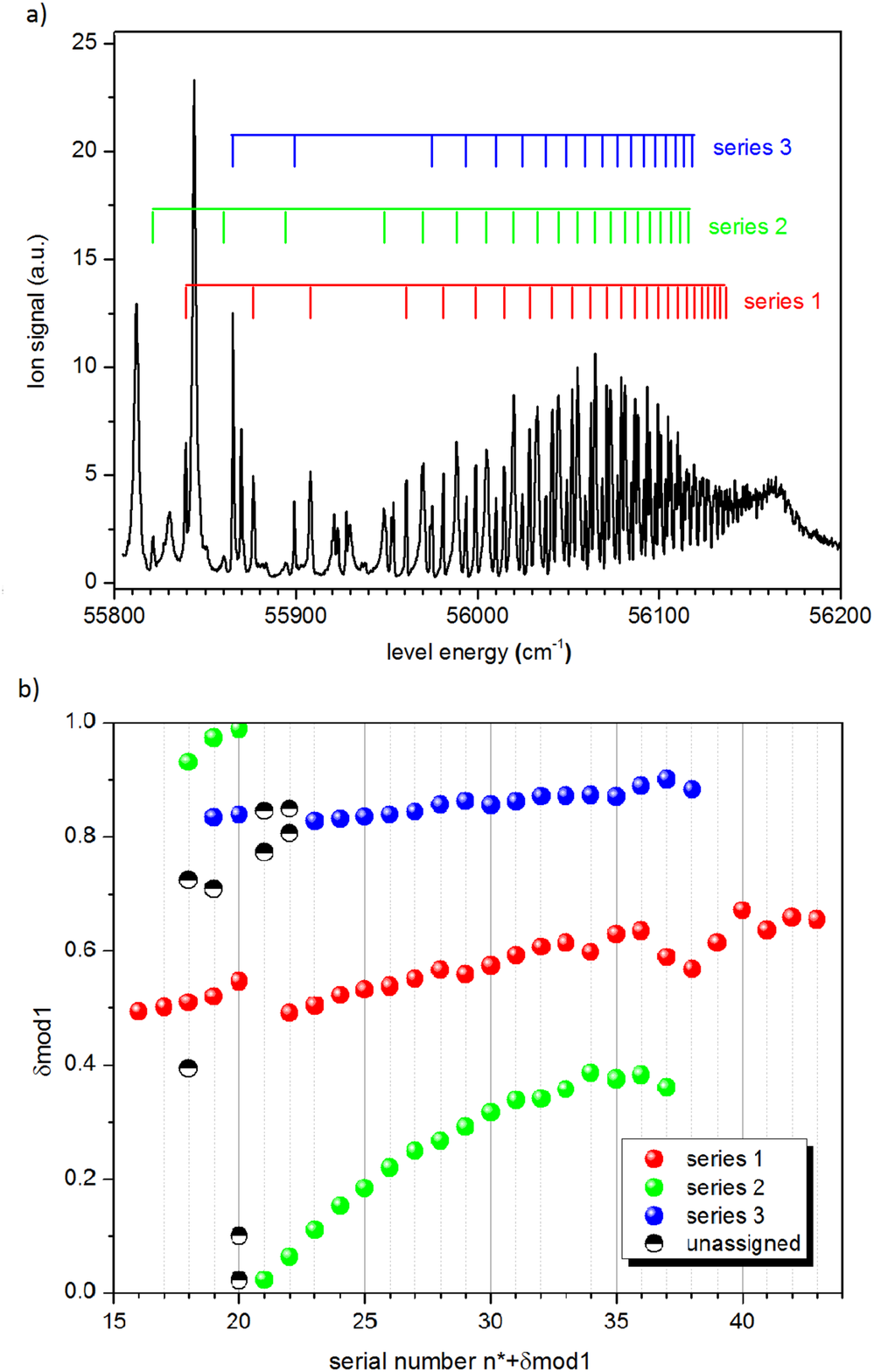}}
\end{center}
\centering
\caption{ a) AI spectrum approaching the limit $5d6s$ $^3D_2$ = 56197.9~cm$^{-1}$. b) $\delta$$mod1$ v.s. serial number ($n$$^*$+$\delta$$mod1$) for three observed AI Rydberg series. A good regularity displays in all three series for high serial number down to the serial number = 23 where perturbations show significant effect.}
\label{2nd_ionic}
\end{figure}

Lu$^+$ has a ground state $6s^2$ $^1S_0$ and three metastable states $5d6s$ $^3D_{1,2,3}$ with the energies of 11796.24~cm$^{-1}$, 12435.32~cm$^{-1}$, 14199.08~cm$^{-1}$ respectively. The relatively low energy of these ionic states gives rise to a rich AI spectrum in region we scanned. In total 340 AI states were observed. Although most of them cannot be classified due to the complexity, six clear AI Rydberg series was clearly observed: three converging to the Lu$^+$ $5d6s$ $^3D_1$ state and three converging to the Lu$^+$ $5d6s$ $^3D_2$ state. The spectra and the corresponding Fano plots are shown in Fig.~\ref{1st_ionic},~\ref{2nd_ionic}.

In the Fano plots the serial number is used as defined as $n$$^*$+$\delta$$mod1$ due to the difficulty in determining the principal number of the AI levels. For the AI states approaching the first ionic state $5d6s$ $^3D_1$, the series 1 and 3 have linewidths scaling with $n$$^*$$^{-3}$ for the serial number 19-30. For higher members, the observed linewidth stays essentially constant around 0.25~cm$^{-1}$. The main contribution is from the laser linewidth ($\sim$7~GHz after frequency doubling). Only series 2 does not show any pronounced change of the linewidth (0.25$\pm$0.10~cm$^{-1}$) across the energy range investigated. Most likely series 2 has a linewidth smaller than our experimental resolution in the investigated range. All three series show significant perturbation around the energy 55450~cm$^{-1}$, which is evident as a broad line intensity dip in Fig.~\ref{1st_ionic}-a and as a rapid variation of the quantum defect in Fig.~\ref{1st_ionic}-b (serial number=32-35). Another strong perturbation happens around 55350~cm$^{-1}$, where the visible regularity ends.

\LTcapwidth=0.5\textwidth
\begin{longtable}{cccccc}
\caption{Even-parity AI Rydberg series converging to Lu$^+$ $5d6s$ $^3D_2$ = 56197.9~cm$^{-1}$ with both the experimental values and the RMCT calculated values in the assigned configurations.}\\
\label{Rydberg_AI_2}
\\
\hline\hline
 &\multicolumn{2}{c}{{\bf experimental}}&& \multicolumn{2}{c}{{\bf RMCT calculation}}\\
serial& \multicolumn{2}{c}{{\bf Series 1 }}&& \multicolumn{2}{c}{5d6s($^3$D$_2$)ns$_{1/2}$}\\
num.& \multicolumn{2}{c}{}&& \multicolumn{2}{c}{J=3/2}\\
\cline{2-3}\cline{5-6}
&	$\sigma$(cm$^{-1}$)	&$\delta$$mod1$& &$\sigma$(cm$^{-1}$)&$\delta$$mod1$\\
\hline
16	&	55741.53	&	0.49	&		&	55750.00	&	0.35	\\
17	&	55794.78	&	0.50	&		&	55801.30	&	0.37	\\
18	&	55839.21	&	0.51	&		&	55843.90	&	0.39	\\
19	&	55876.59	&	0.52	&		&	55886.90	&	0.22	\\
20	&	55907.93	&	0.55	&		&	55914.20	&	0.33	\\
21	&		&		&		&	55939.90	&	0.38	\\
22	&	55960.72	&	0.49	&		&	55957.36	&	0.64	\\
23	&	55981.08	&	0.50	&		&	55982.32	&	0.44	\\
24	&	55998.83	&	0.52	&		&	55999.63	&	0.48	\\
25	&	56014.63	&	0.53	&		&	56014.98	&	0.51	\\
26	&	56028.65	&	0.54	&		&	56028.66	&	0.54	\\
27	&	56041.05	&	0.55	&		&	56040.91	&	0.56	\\
28	&	56052.11	&	0.57	&		&	56051.91	&	0.59	\\
29	&	56062.25	&	0.56	&		&	56061.83	&	0.60	\\
30	&	56071.19	&	0.57	&		&	56070.78	&	0.62	\\
31	&	56079.24	&	0.59	&		&	56078.90	&	0.64	\\
32	&	56086.57	&	0.61	&		&	56086.27	&	0.65	\\
33	&	56093.29	&	0.61	&		&	56092.99	&	0.66	\\
34	&	56099.56	&	0.60	&		&	56099.13	&	0.67	\\
35	&	56105.03	&	0.63	&		&	56104.75	&	0.68	\\
36	&	56110.18	&	0.64	&		&	56109.91	&	0.69	\\
37	&	56115.15	&	0.59	&		&	56114.66	&	0.70	\\
38	&	56119.60	&	0.57	&		&	56119.04	&	0.70	\\
39	&	56123.44	&	0.61	&		&	56123.08	&	0.71	\\
40	&	56126.97	&	0.67	&		&	56126.82	&	0.71	\\
41	&	56130.56	&	0.64	&		&	56130.29	&	0.72	\\
42	&	56133.71	&	0.66	&		&	56133.51	&	0.72	\\
43	&	56136.72	&	0.65	&		&	56136.51	&	0.73	\\
\hline\hline
\end{longtable}

\begin{ThreePartTable}
\begin{TableNotes}
\footnotesize
 \item[*] same serial number for two resonances due to a jump in order.
 \item[a] asymmetric resonance peaks.
\end{TableNotes}
\begin{longtable}{cccccc}
\hline\hline
 &\multicolumn{2}{c}{{\bf experimental}}&& \multicolumn{2}{c}{{\bf RMCT calculation}}\\
serial& \multicolumn{2}{c}{{\bf Series 2 }}&& \multicolumn{2}{c}{5d6s($^3$D$_2$)nd$_{3/2}$}\\
num.& \multicolumn{2}{c}{}&& \multicolumn{2}{c}{J=3/2}\\
\cline{2-3}\cline{5-6}
&	$\sigma$(cm$^{-1}$)	&$\delta$$mod1$& &$\sigma$(cm$^{-1}$)&$\delta$$mod1$\\
\hline
18	&	55821.26		&	0.93	&		&	55832.20	&	0.68	\\
19	&	55860.20		&	0.97	&		&	55866.57	&	0.80	\\
20	&	55894.29		&	0.99	&		&	55898.59	&	0.85	\\
21\tnote{*}	&			&		&		&	55926.40	&	0.90	\\
21\tnote{*}	&	55948.53\tnote{a}	&	0.02	&		&	55950.46	&	0.94	\\
22	&	55969.87\tnote{a}	&	0.06	&		&	55971.34	&	0.99	\\
23	&	55988.46\tnote{a}	&	0.11	&		&	55989.64	&	0.05	\\
24	&	56004.95		&	0.15	&		&	56005.85	&	0.10	\\
25	&	56019.73		&	0.18	&		&	56020.16	&	0.15	\\
26	&	56032.80		&	0.22	&		&	56033.10	&	0.20	\\
27	&	56044.58		&	0.25	&		&	56044.78	&	0.23	\\
28	&	56055.24		&	0.27	&		&	56055.36	&	0.26	\\
29	&	56064.77		&	0.29	&		&	56064.99	&	0.27	\\
30	&	56073.38		&	0.32	&		&	56073.57	&	0.29	\\
31	&	56081.20		&	0.34	&		&	56081.45	&	0.30	\\
32	&	56088.43		&	0.34	&		&	56088.55	&	0.32	\\
33	&	56094.93		&	0.36	&		&	56095.15	&	0.32	\\
34	&	56100.80		&	0.39	&		&	56101.12	&	0.33	\\
35	&	56106.39		&	0.38	&		&	56106.53	&	0.35	\\
36	&	56111.42		&	0.38	&		&	56111.61	&	0.34	\\
37	&	56116.18		&	0.36	&		&	56116.24	&	0.35	\\
\hline\hline
\insertTableNotes
\end{longtable}
\end{ThreePartTable}

\begin{longtable}{ccccccccc}
\hline\hline
& \multicolumn{2}{c}{{\bf experimental}}&& \multicolumn{5}{c}{{\bf RMCT calculation}}\\
serial& \multicolumn{2}{c}{{\bf Series 3 }}&& \multicolumn{2}{c}{5d6s($^3$D$_2$)nd$_{3/2}$}&&\multicolumn{2}{c}{5d6s($^3$D$_2$)nd$_{5/2}$}\\
num.& \multicolumn{2}{c}{}&& \multicolumn{2}{c}{J=1/2}&&\multicolumn{2}{c}{J=5/2}\\
\cline{2-3}\cline{5-6}\cline{8-9}
&	$\sigma$(cm$^{-1}$)	&$\delta$$mod1$& &$\sigma$(cm$^{-1}$)&$\delta$$mod1$& &$\sigma$(cm$^{-1}$)&$\delta$$mod1$\\
\hline
19	&	55865.38	&	0.83	&		&	55839.05	&	0.51	&		&	55862.60	&	0.91	\\
20	&	55899.04	&	0.84	&		&	55899.17	&	0.83	&		&	55907.74	&	0.55	\\
21	&		&		&		&	55927.64	&	0.85	&		&		&		\\
22	&		&		&		&	55952.40	&	0.86	&		&		&		\\
23	&	55974.71	&	0.83	&		&	55973.91	&	0.87	&		&	55977.26	&	0.70	\\
24	&	55993.49	&	0.83	&		&	55992.84	&	0.87	&		&	55995.26	&	0.73	\\
25	&	56010.00	&	0.83	&		&	56009.35	&	0.88	&		&	56011.16	&	0.76	\\
26	&	56024.58	&	0.84	&		&	56023.98	&	0.88	&		&	56025.28	&	0.79	\\
27	&	56037.51	&	0.84	&		&	56037.03	&	0.88	&		&	56037.87	&	0.82	\\
28	&	56048.98	&	0.86	&		&	56048.66	&	0.89	&		&	56049.16	&	0.84	\\
29	&	56059.31	&	0.86	&		&	56059.06	&	0.89	&		&	56059.29	&	0.86	\\
30	&	56068.72	&	0.86	&		&	56068.37	&	0.90	&		&	56068.45	&	0.89	\\
31	&	56077.10	&	0.86	&		&	56076.82	&	0.90	&		&	56076.72	&	0.91	\\
32	&	56084.67	&	0.87	&		&	56084.47	&	0.90	&		&	56084.23	&	0.93	\\
33	&	56091.61	&	0.87	&		&	56091.43	&	0.90	&		&	56091.06	&	0.95	\\
34	&	56097.92	&	0.87	&		&	56097.76	&	0.90	&		&	56097.29	&	0.98	\\
35	&	56103.71	&	0.87	&		&	56103.53	&	0.90	&		&	56102.98	&	0.00	\\
36	&	56108.90	&	0.89	&		&	56108.81	&	0.91	&		&	56108.19	&	0.03	\\
37	&	56113.71	&	0.90	&		&	56113.66	&	0.91	&		&	56112.98	&	0.06	\\
38	&	56118.27	&	0.88	&		&	56118.14	&	0.91	&		&	56117.40	&	0.08	\\
\hline\hline
\end{longtable}
\addtocounter{table}{-2}

For the AI states approaching the second ionic state $5d6s$ $^3D_2$, a clear regularity displays in the spectrum until reaching the low energy end around 55930~cm$^{-1}$ (Fig.~\ref{2nd_ionic}). Three  series are distinctive for serial numbers $\geq$23. The linewidth of these AI states are larger than those of the AI states approaching the first ionic state. This is in part due to the increased laser linewidth ($\sim$14~GHz after frequency doubling) in this wavelength range. The linewidth decrease with $n$$^*$$^{-3}$ for serial number 23-30 in series 2 and start to approach the laser resolution $\sim$0.45~cm$^{-1}$ after that. For series 1 and 3, the linewidths stay constant at 0.45$\pm$0.15~cm$^{-1}$, which implies their linewidths well below the laser resolution.

\section{Relativistic multichannel theory (RMCT) calculations}\label{RMCT}

To clearly classify energy levels and better understand the observed experimental spectra, we have performed theoretical calculations on Lu using relativistic multichannel theory (RMCT) within the framework of multichannel quantum defect theory (MQDT) \cite{fano70,lee73,Fano,gree79,lee80,ljm80,seaton83,ljm83}. In MQDT the wavefunction of a discrete state in a Coulomb potential (such as Rydberg states) can be described as a superposition of the wavefunctions of a group of dissociation channels. The coefficients of the superposition depend on the interactions between the channels, which are characterized by a set of physical MQDT parameters ($\mu$$_{\alpha}$, $U_{i\alpha}$ ). Both the discrete states near a threshold and the adjacent continuum can be treated in an unified manner with the MQDT parameters, which makes it well suitable to analyze Rydberg and AI states. 

\begin{table}[!htbp] \footnotesize
\caption{Lu dissociation channels included in the RMCT calculation.}
\begin{center}
\begin{threeparttable}
\begin{tabular}{ccc}
\hline\hline
$J^{\pi}=(1/2)^{+}$                      &   $J^{\pi}=(3/2)^{+}$                 & $J^{\pi}=(5/2)^{+}$                    \\
\hline                                                                                                                    \\
    $6s^2(^1S_0)\varepsilon d_{3/2}$           &     $6s^2(^1S_0)\varepsilon d_{3/2}$                   &   $6s^2(^1S_0)\varepsilon d_{5/2}$   \\
    $5d6s(^3D_1)\varepsilon s_{1/2}$           &     $5d6s(^3D_1)\varepsilon s_{1/2}$                   &   $5d6s(^3D_1)\varepsilon d_{3/2}$   \\
    $5d6s(^3D_1)\varepsilon d_{3/2}$           &     $5d6s(^3D_1)\varepsilon d_{3/2}$                   &   $5d6s(^3D_1)\varepsilon d_{5/2}$   \\
    $5d6s(^3D_2)\varepsilon d_{3/2}$           &     $5d6s(^3D_1)\varepsilon d_{5/2}$                   &   $5d6s(^3D_2)\varepsilon s_{1/2}$   \\
    $5d6s(^3D_2)\varepsilon d_{5/2}$           &     $5d6s(^3D_2)\varepsilon s_{1/2}$                   &   $5d6s(^3D_2)\varepsilon d_{3/2}$   \\
    $5d6s(^3D_3)\varepsilon d_{5/2}$           &     $5d6s(^3D_2)\varepsilon d_{3/2}$                   &   $5d6s(^3D_2)\varepsilon d_{5/2}$   \\
    $5d6s(^1D_2)\varepsilon d_{3/2}$           &     $5d6s(^3D_2)\varepsilon d_{5/2}$                   &   $5d6s(^3D_3)\varepsilon s_{1/2}$   \\
    $5d6s(^1D_2)\varepsilon d_{5/2}$           &     $5d6s(^3D_3)\varepsilon d_{3/2}$                   &   $5d6s(^3D_3)\varepsilon d_{3/2}$   \\
    $6s6p(^3P_0)\varepsilon p_{1/2}$           &     $5d6s(^3D_3)\varepsilon d_{5/2}$                   &   $5d6s(^3D_3)\varepsilon d_{5/2}$   \\
    $6s6p(^3P_1)\varepsilon p_{1/2}$           &     $5d6s(^1D_1)\varepsilon s_{1/2}$                   &   $5d6s(^1D_2)\varepsilon s_{1/2}$   \\
    $6s6p(^3P_1)\varepsilon p_{3/2}$           &     $5d6s(^1D_1)\varepsilon d_{3/2}$                   &   $5d6s(^1D_2)\varepsilon d_{3/2}$   \\
    $6s6p(^3P_2)\varepsilon p_{3/2}$           &     $5d6s(^1D_1)\varepsilon d_{5/2}$                   &   $5d6s(^1D_2)\varepsilon d_{5/2}$   \\
    $6s6p(^3P_2)\varepsilon f_{5/2}$           &     $6s6p(^3P_0)\varepsilon p_{3/2}$                   &   $6s6p(^3P_0)\varepsilon f_{5/2}$   \\
    $6s6p(^1P_1)\varepsilon p_{1/2}$           &     $6s6p(^3P_1)\varepsilon p_{1/2}$                   &   $6s6p(^3P_1)\varepsilon p_{3/2}$   \\
    $6s6p(^1P_1)\varepsilon p_{3/2}$           &     $6s6p(^3P_1)\varepsilon p_{3/2}$                   &   $6s6p(^3P_1)\varepsilon f_{5/2}$   \\
                                               &     $6s6p(^3P_2)\varepsilon p_{1/2}$                   &   $6s6p(^3P_2)\varepsilon p_{1/2}$   \\
                                               &     $6s6p(^3P_2)\varepsilon p_{3/2}$                   &   $6s6p(^3P_2)\varepsilon p_{3/2}$   \\
                                               &     $6s6p(^3P_2)\varepsilon f_{5/2}$                   &   $6s6p(^3P_2)\varepsilon f_{5/2}$   \\
                                               &     $6s6p(^3P_2)\varepsilon f_{7/2}$                   &   $6s6p(^3P_2)\varepsilon f_{7/2}$   \\
                                               &     $6s6p(^1P_1)\varepsilon p_{1/2}$                   &   $6s6p(^1P_1)\varepsilon p_{3/2}$   \\
                                               &     $6s6p(^1P_1)\varepsilon p_{3/2}$                   &   $6s6p(^1P_1)\varepsilon f_{5/2}$   \\
                                               &     $6s6p(^1P_1)\varepsilon f_{5/2}$                   &   $6s6p(^1P_1)\varepsilon f_{7/2}$  \\
\hline\hline
\end{tabular}
   \end{threeparttable}
\end{center}
\label{dis_channel}
\end{table}

\begin{figure*}[!htbp]
\begin{center}
\centerline{\includegraphics[width=0.9\textwidth]{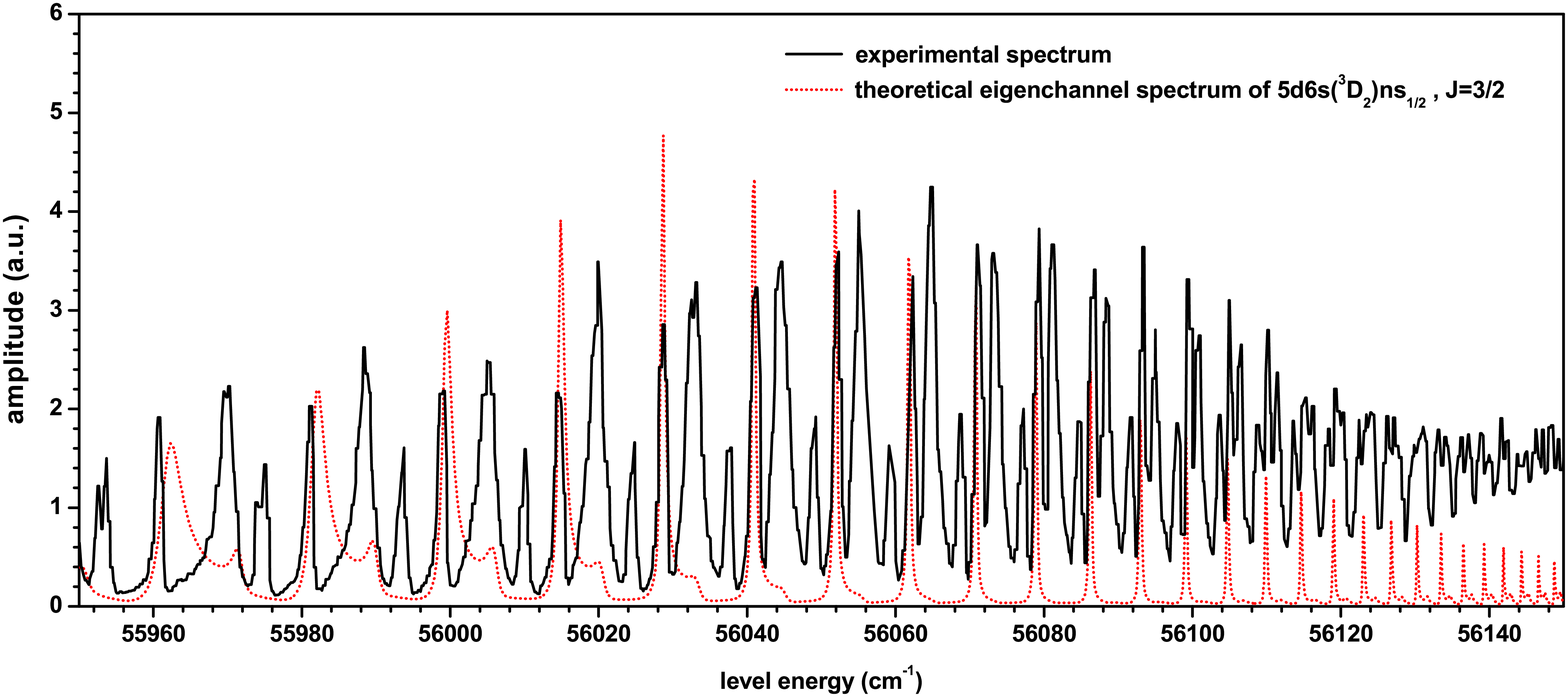}}
\end{center}
\centering
\caption{Comparison of the experimental spectrum with the RMCT calculated eigenchannel spectrum. The presented experimental spectrum is converging to Lu$^+$ 5d6s $^3$D$_2$ state. The eigenchannel spectrum shown is for the $5d6s$($^3D_2$)$ns$$_{1/2}$ J=3/2 seres.}
\label{example}
\end{figure*}

For a small number of channels, the MQDT parameters can be semiempirically obtained by fitting spectroscopic data. However with the increasing number of the channels involved, this method is hampered by complicated and laborious numerical fitting with too many parameters and the need of complete spectroscopic data. Another approach is to calculate the MQDT parameters directly from first principles with relativistic multichannel theory (RMCT)\cite{lee74,lee89,zou95,huang95,yan96,xia01,xia03}, which can be regarded as an extension of the traditional configuration interaction (CI) method by including the continuum. It has been successfully employed to calculate the Rydberg and autoionization Rydberg spectrum of scandium with three valence electrons. The calculated spectrum was in general agreement with the experimental spectrum as a whole, and the Rydberg states and autoionization states were assigned through comparing the calculated eigenchannel spectrum with the experimental spectra\cite{Jia09}. 

In this work, the MQDT parameters are calculated by RMCT firstly at some chosen energy points in the energy range investigated. Because the MQDT parameters ($\mu_{\alpha}$,$U_{i\alpha}$) are smoothly energy dependent within the neighborhood of ionization limits\cite{Fano}, the MQDT parameters at any energy can be easily obtained using interpolation or extrapolation. The experimental spectrum of Lu was obtained via the intermediate state $5d6s6p(^1D)$ $^2D^{\circ}_{3/2}$, therefore the channels in $J^{\pi}=(1/2)^+,(3/2)^+,(5/2)^+$ symmetry should be considered in the calculations based on the selection rules. The dissociation channels included in the RMCT calculation are listed in Tab.~\ref{dis_channel}. Nine ionic states with the configurations of 6s$^2$, 5d6s and 6s6p were considered. 

For the energy range of the Rydberg states, all the dissociation channels are closed and the level energies can be directly calculated. The calculated level energies are presented with the experimental data in Tab.~\ref{table Rydberg_D_3/2}-\ref{table Rydberg_S_1/2} for comparison. A good agreement is found within 1~cm$^{-1}$ for n$\geq$22 of $6s^2nd$ $^2D_{3/2}$, n$\geq$38 of $6s^2nd$ $^2D_{5/2}$ and n$\geq$26 of $6s^2ns$ $^2S_{1/2}$.

Different from Rydberg series below the IP, AI series can be a superposition of multiple channels. The assignments of AI states in this work was made by comparing the experimental spectrum with different eigenchannel spectra in terms of energy position. The eigenchannel spectra were calculated by setting $D_{\alpha'}=\delta_{\alpha \alpha'}$ for the eigenchannel $\alpha$\cite{Jia09}. A example of the comparison is shown in Fig.~\ref{example}. Clearly the eigenchannel spectrum can not completely reflect the profile/position of the experimentally observed resonant peaks. For some strongly perturbed resonances this method may even fail. However it generally meets the purpose of assignment in most cases\cite{Jia09}. To avoid confusion, in this work only the clearly grouped AI states were assigned.

Due to channel interaction, the calculated eigenchannel spectral peaks sometimes present a Fano profile or even irregular shapes like Shore profile \cite{Shore}. For the simplicity and the unification of the treatment, a maximum center was employed as a reasonable approximation by providing enough calculated data points on the peak profile. The summary of the RMCT theoretical level energies with the assigned configurations are presented in Tab.~\ref{Rydberg_AI_1} and \ref{Rydberg_AI_2} to compare with the experimental results. In some cases more than one eigenchannel spectrum is consistent with an observed AI series in terms of energy position, which implies strong configuration interaction between these eigenchannel wavefunctions and mixed components of the observed AI series. To avoid omission, all possible assignments are listed in the tables.

\section{Conclusion}\label{Conclusion}

Even-parity Rydberg and AI states of Lu were studied by means of laser resonance ionization spectroscopy. Three Rydberg series $nd$ $^2D_{3/2}$, $nd$ $^2D_{5/2}$ and $ns$ $^2S_{1/2}$ converging to the IP were measured by means of laser resonance ionization spectroscopy, and the spectrum was interpreted. In addition six autoionizing (AI) series converging to the core states of $5d6s$ $^3D_1$ and $5d6s$ $^3D_2$ were observed and reported for the first time. The assignment of measured AI series was attempted with the aid of RMCT calculations. Experimental values for both Rydberg and AI states have been compared with the RMCT calculations. For the Rydberg states the agreement is within 1~cm$^{-1}$ for high n members ($>$22 for $6s^2nd$ $^2D_{3/2}$). The comparison of the experimental level energies and the RMCT calculation with assigned configurations for AI states is also presented. 

\begin{acknowledgments}
The experimental work has been funded by TRIUMF which receives federal funding via a contribution agreement with the National Research Council of Canada and through a Natural Sciences and Engineering Research Council of Canada (NSERC) Discovery Grant (386343-2011). The theoretical work is supported by Natural Science Foundation of China (11604334) and Beijing Natural Science Foundation (1164016). M. Mostamand acknowledges funding through the University of Manitoba graduate fellowship. 
\end{acknowledgments}









\bibliographystyle{elsarticle-num}

\end{document}